\documentclass[oupdraft]{bio}
\usepackage{url}

\usepackage{amssymb, amsmath, amsfonts, mathrsfs, bbm}
\usepackage{xcolor}
\usepackage{multirow}
\usepackage{siunitx, booktabs,threeparttable}
\usepackage{nicefrac}

\newcommand{\prob}{\mathbb{P}}
\newcommand{\var}{\textnormal{Var}}
\newcommand{\ind}{\mathbbm{1}}
\newcommand{\E}{\mathbb{E}}

\newcommand\independent{\protect\mathpalette{\protect\independenT}{\perp}}
\def\independenT#1#2{\mathrel{\rlap{$#1#2$}\mkern2mu{#1#2}}}

\definecolor{amber}{rgb}{1.0, 0.49, 0.0}

\newtheorem{assumption}{Assumption}

\begin{document}

\title{A Generalizability Score for Aggregate Causal Effect}

\author{Rui Chen$^1$, Guanhua Chen$^{2,\ast}$, Menggang Yu$^{2,\ast}$\\[14pt]
\textit{
$^1$Department of Statistics, University of Wisconsin, Madison, WI, U.S.A.\\
$^2$Department of Biostatistics and Medical Informatics, University of Wisconsin, Madison, WI, U.S.A.}
\\[2pt]
{gchen25@wisc.edu; meyu@biostat.wisc.edu}}

\markboth%
{R. Chen, G. Chen and M. Yu}
{Generalizability Score}

\maketitle

\footnotetext{To whom correspondence should be addressed.}

\begin{abstract}
{ Scientists frequently generalize population level causal quantities such as average treatment effect from a source population to a target population. When the causal effects are heterogeneous, differences in subject characteristics between the source and target populations may make such a generalization difficult and unreliable. Reweighting or regression can be used to adjust for such differences when generalizing. However, these methods typically suffer from large variance if there is limited covariate distribution overlap between the two populations. We propose a generalizability score to address this issue. The score can be used as a yardstick to select target subpopulations for generalization. A simplified version of the score avoids using any outcome information and thus can prevent deliberate biases associated with inadvertent access to such information. Both simulation studies and real data analysis demonstrate convincing results for such selection.
}
{Average treatment effect; Generalizability; Propensity score; Treatment effect heterogeneity.}
\end{abstract}


\section{Introduction}
\label{sec:intro}

Scientists routinely make causal generalization in their research. This perplexing scientific and philosophical problem demands resolution of many challenging issues \citep{Shadishbook} when the causal effect is possibly heterogeneous or depends on subject characteristics. Such generalizability \citep{Cole2010,tipton2013improving,buchanan2018generalizing} is also known as external validity \citep{Rothwell2005}, or transportability \citep{pearl2014external,bareinboim2016causal,Rudolph2017} in literature. In some recent papers \citep{dahabreh2020extending,degtiar2021review} terms ``generalizability'' and ``transportability'' bear different meanings. Our paper focuses on the setting where the source population is external to the target population.

For average treatment effect (ATE) estimation, properly planned and conducted randomized trials are internally valid; however, not necessarily generalizable in the presence of heterogeneous treatment effect. In other words, the unbiased estimate of the ATE from a randomized trial may not equal to the ATE of a target population if trial participants can not well represent the target population with respect to the distribution of effect modifiers \citep{dahabreh2020extending}.
While exploring such heterogeneity is itself of great interest, this article focuses on generalizing aggregate causal quantities such as the ATE. 

In the past decade, an active area of such causal generalization research is to bridge findings from a randomized trial to a target population \citep{Cole2010,tipton2013improving,Rudolph2017,buchanan2018generalizing,dahabreh2020extending}. Most of these methods rely on modeling the trial participation probability, which quantifies the similarity between trial participants and patients in the target population. The estimated probability is used in the subsequent analysis for reweighting \citep{Cole2010,buchanan2018generalizing} or post-stratification \citep{Cole2010}. Some existing methods also incorporate outcome modeling to improve estimation efficiency, such as the targeted maximum likelihood estimators \citep{Rudolph2017} and augmented inverse probability weighted estimators \citep{dahabreh2020extending,yang2020doubly}.

Although these approaches effectively adjust for the compositional difference between the source participants and the target population and reduce estimation bias, an essential premise is overlap, which essentially requires that every individual in the target population have matched source participants with similar characteristics. When there is insufficient overlap, reweighting-based adjustment usually introduces large variability to the estimation result, and outcome modeling approaches rely heavily on extrapolation and thus are also unstable.

We consider this same research problem of causal generalization from a source population to a target population, although we assume that treatments may have been given to the source population in a possibly non-randomized fashion as in typical observational studies. The corresponding outcomes and subject characteristics or covariates have been observed in the source population. On the other hand, only subject characteristics have been collected in the target population. Our goal is also to characterize the overlap between the two populations on the basis of observed characteristics so that the generalization is most stable. \citet{stuart2011use} proposed to assess the similarity between a cohort of trial participants and a target population using the difference in the mean participation probability; \citet{tipton2014generalizable} adopted a similar strategy, but replaced the difference in mean with a distributional difference. However, these works did not quantify how the proposed metrics relate to the causal effect estimation, or provide remedies to cope with insufficient overlap. Further, their methods are limited to the scenarios where the source population data is from randomized trials.

The issue of limited overlap has been studied by many researchers in the context of observational studies for a single population, where estimation relies on sufficient overlap between the treatment arms. The ``overlap'' is then defined in terms of propensity score or the probability of treatment assignment. A common approach is to restrict the population of interest to a subset that has sufficient overlap. \citet{dehejia1999causal} and \citet{lopez2017estimation} discarded individuals with very small or large propensity scores. \citet{crump2009dealing} justified this approach from a semiparametric efficiency perspective and suggested rule-of-thumb propensity score thresholds of 0.1 and 0.9 to trim the population. \citet{crump2009dealing}'s approach was then extended to multiple treatment cases \citep{yang2016propensity}. Another popular approach is to find a weighted ATE that is least affected by limited overlap \citep{li2018balancing}.

All the above methods dealing with limited overlap boil down to defining a new estimand for aggregate causal effect estimation, either by subsetting or reweighting the study population, in a data-dependent manner. An immediate question is the implication of the resulting estimand. This prompted \cite{Rosenbaum.jcgs.2011} to introduce the concept of marginal subjects as those having some probability of receiving the treatment (i.e. with sufficient overlap). The estimand is then the aggregate causal effect for the subpopulation of marginal subjects. To enhance interpretability of such subpopulation, \cite{traskin2011defining} developed a tree approach.

This paper focuses on generalizing causal estimands from a source population to a target population with potentially limited covariate overlap. We address limited overlap in terms of participation probability and propensity score simultaneously by characterizing their impact on estimation precision based on the semiparametric efficiency framework \citep{tsiatis2007semiparametric}. A key quantity, which will be referred to as the \textit{generalizability score}, is then introduced as a yardstick to evaluate and select subpopulations of the target population for causal generalization. Selection based on the score results in optimal efficiency of causal generalization among all subsets that cover the same proportion of the target population. A plot of the generalizability score also enables evaluating the sensitivity of the estimation efficiency to different proportions of the target population and therefore facilitates practical choices for generalization. A simplified version of the score avoids using any outcome information from the source population, and thus can prevent introducing deliberate biases associated with inadvertent access to such information \citep{crump2006moving,Rubinsim2007,rubin2008}. Both simulation studies and real data analysis demonstrate convincing results for such selection. Because our selection of the subset can be done without accessing outcome data, the logic of existing approaches to deal with the definition of the resulting estimands \citep{crump2009dealing,traskin2011defining,Rosenbaum.jcgs.2011} is applicable to our paper. 

In Section \ref{sec:prelim}, we introduce the framework and underlying assumptions, followed by a more detailed discussion on the impact of limited overlap. In Section \ref{sec:method}, we present the major methodological results, where we derive a semiparametric efficiency bound for the estimation task. The efficiency bound naturally gives rise to the notion of the generalizability score. We illustrate the proposed approach through simulation studies and a real data example in Section \ref{sec:simu} and \ref{sec:data}. We conclude the paper with a discussion in Section \ref{sec:disc}.


\section{Preliminaries}
\label{sec:prelim}
\subsection{Notations and assumptions}
\label{sec:framework}

Suppose we observe two samples from two distinct populations $\mathcal{S}$ (source) and $\mathcal{T}$ (target). The source sample is from a randomized trial or an observational study and for subject $i$ we observe covariate $X_i \in \mathcal{X}$, treatment assignment $A_i\in\{0, 1\}$, and a corresponding outcome $Y_i$. For individuals in the target sample, we only have their covariate information $X_i$.  Let $S_i$ be the population indicator such that $S_i=1$ for $i\in {\cal S}$ and $S_i=0$ for $i\in {\cal T}$. Therefore our observed data consist of $\{ (X_i, A_i, Y_i, S_i)\!:\, S_i=1 \}$ and $\{ (X_i, S_i)\!:\,  S_i=0 \}$. The total sample size is $n$.

We use the potential outcome framework \citep{rosenbaum1983central,imbens_rubin_2015} to formulate the causal problem. Under the Stable Unit Treatment Value Assumption (SUTVA), which posits no interference between different individuals and no hidden variation of treatments, each individual $i$ has two potential outcomes $Y_i(0)$ and $Y_i(1)$, the values of the outcome that would be observed if $i$ were to receive control or treatment, respectively. Then the observed outcome in the source sample is $Y_i = Y_i(A_i)$. We associate each observation with a ``full'' random variate $(X_i , S_i , A_i , Y_i(0), Y_i(1))$, which across $i$ are assumed to be i.i.d. draws from a joint distribution of $(X, S, A, Y(0), Y(1))$. All the probabilities and expected values below are taken with respect to this distribution.

We assume that the treatment assignment mechanism in the source sample is determined by a propensity score $\pi(x) = \prob (A = 1 \mid X = x, S=1)$ \citep{rosenbaum1983central}.  If the source sample is from a randomized trial, then $\pi(x)$ is known. In general, it needs to be estimated. We further denote $\rho(x) = \prob(S=1 \mid X=x)$ and refer this as participation probability \citep{dahabreh2020extending}.

To identify the causal effect for the target population, we will work with the following four assumptions in addition to the SUTVA. The first two, adopted from \citet{rosenbaum1983central}, are for identification of aggregate causal effects on the source population. Assumptions 3 and 4, adopted from \citet{Rudolph2017} and \citet{Dahabreh2019}, link the source population $\mathcal{S}$ to the target population $\mathcal{T}$ and enable us to generalize the aggregate causal effects.  

\begin{assumption}[Unconfoundedness of treatment assignment]
\label{assp:unconf}
    In the source population, $(Y(0), Y(1))$ are conditionally independent of $A$ given $X$:
    $(Y(0), Y(1)) \independent A \mid X, S=1$.
\end{assumption}
\begin{assumption}[Overlap of propensity score]
\label{assp:ovlptrt}
    The propensity score of the source population is bounded away from 0 and 1: for some $c > 0$, $c \le \pi(X) \le 1 - c$ almost surely. 
\end{assumption}

\begin{assumption}[Causal exchangability between populations]
\label{assp:exchange}
    The source population and the target population have the same conditional average treatment effect (CATE): $\E \{Y(1) - Y(0) \mid X, S=0\} = \E \{Y(1) - Y(0) \mid X, S=1\}$ almost surely.
\end{assumption}
\begin{assumption}[Overlap of participation probability]
\label{assp:ovlppop}
    Conditional on the covariates, the participation probability is bounded away from 0: $\rho(X) > c$ almost surely for some $c > 0$. 
\end{assumption}

Assumptions \ref{assp:ovlptrt} and \ref{assp:ovlppop} require the propensity score be bounded away from 0 and 1 and the participation probability be bounded away from 0. This is to ensure overlap in the covariate distributions. However, if either of them only holds for very small value of $c$, the overlap might not be sufficient to guarantee stable estimation. We will elucidate this issue in the next section and the focus of this paper is to deal with it. Note that propensity score and participation probability could depend on different sets of covariates. In particular, the propensity score only depends on the confounders, and the participation probability only depends on the effect modifiers \citep{stuart2011use}. However, to make such a distinction, it would require strong prior knowledge related to the mechanism of treatment assignment and study participation \citep{vanderweele2012confounding}. Therefore, for conciseness of presentation, we use a unified symbol $X$ to denote all the covariates, not dividing them into different sets.

We denote the conditional mean and variance of the potential outcomes in the source population by 
$\mu_{a, \mathcal{S}}(x) = \E\{Y(a)\mid X=x, S=1\},
\sigma^2_{a, \mathcal{S}}(x) = \var\{ Y(a)\mid X=x, S=1\}.$

One can also define $\mu_{a, \mathcal{T}}(x)$ and $\sigma^2_{a, \mathcal{T}}(x)$ similarly. The CATE for the source population is $\theta_\mathcal{S}(x) =\mu_{1, \mathcal{S}}(x) - \mu_{0, \mathcal{S}}(x)$ and for the target population is $\theta_\mathcal{T}(x) =\mu_{1, \mathcal{T}}(x) - \mu_{0, \mathcal{T}}(x)$. Assumption \ref{assp:exchange} states that there is a common CATE $\theta(x)$:
\begin{equation*}
\theta(x) =\theta_\mathcal{S}(x) =\theta_\mathcal{T}(x). 
\end{equation*}

The ATEs for the source and target populations are $$\theta_\mathcal{S} = \E\big\{\theta(X) \mid S = 1 \big\}, \quad  \theta_\mathcal{T} = \E\{ \theta(X) \mid S = 0\}.$$
Proof of identifiability for our estimand $\theta_{\cal T}$ under the aforementioned assumptions is provided in the Supplementary Materials.
Note that even though $\theta_\mathcal{S}(x) =\theta_\mathcal{T}(x)$ under Assumption \ref{assp:exchange},  we generally have $\theta_\mathcal{S} \not=\theta_\mathcal{T}$ as the distributions of $X$ may differ between $\mathcal{S}$  and $\mathcal{T}$ unless $\rho(x)$ is a constant.

\subsection{Impact of limited overlap}
\label{sec:impact_ovlp}

We first recap the impact of limited overlap on estimating $\theta_\mathcal{S}$ in observational studies. Under the SUTVA and Assumptions \ref{assp:unconf} and \ref{assp:ovlptrt}, many estimators have been developed to consistently estimate $\theta_\mathcal{S}$ from the observed data, such as the inverse probability weighted (IPW) estimators \citep{lunceford2004stratification} or outcome regression estimators \citep{hahn1998role}. The overlap in the treatment arms plays a crucial role in the stable estimation of $\theta_\mathcal{S}$ as it ensures that there are comparable subjects in different treatment arms. For example, the IPW estimators proceed by reweighting each observation with the inverse of the probability of getting the assigned treatment, so it might put extremely large weights on a few observations in the presence of limited overlap as the propensity score approaches to 0 or 1 for some covariate values. Consequently, the IPW estimators will be unduly influenced by these extreme weights. Similarly, the outcome regression estimators will also suffer from high instability in such scenarios because they essentially rely on imputing the potential outcomes, which might unreliable on the region with limited overlap. A common approach to alleviate this issue is to restrict the attention to areas of data with sufficient overlap \citep{dehejia1999causal,crump2009dealing}. 

The problem of insufficient overlap is exacerbated for the estimation of $\theta_\mathcal{T}$. In order to generalize the source sample information to estimate $\theta_\mathcal{T}$, we need to address both the covariate overlap between the treatment arms in the source population, and the covariate overlap between the two populations. That is, stable estimation of $\theta_\mathcal{T}$ relies on finding similar source samples (from both arms) for each individual in the target sample.

To see this, let us consider a H\'ajek-type IPW estimator, which is a natural reweighting method to adjust for compositional difference \citep{Cole2010,dahabreh2020extending}:
\begin{equation}
\label{eq:Hajek}
    \hat{\theta}_{\mathcal{T}}^{(IPW)} =
    \frac{\sum_{i=1}^n \hat{w}_{i1} S_i A_i Y_i}
         {\sum_{i=1}^n \hat{w}_{i1} S_i A_i} 
    -
    \frac{\sum_{i=1}^n \hat{w}_{i0} S_i (1-A_i) Y_i}
         {\sum_{i=1}^n \hat{w}_{i0} S_i (1-A_i)} 
,\end{equation}
where
\begin{equation}
\label{eq:weights}
    \hat{w}_{i1} = \frac{1-\hat{\rho}(X_i)}{\hat{\rho}(X_i)\hat{\pi}(X_i)}  
    \text{ and }
    \hat{w}_{i0} = \frac{1-\hat{\rho}(X_i)}{ \hat{\rho}(X_i) \{1 - \hat{\pi}(X_i)\}}
,\end{equation}
and $\hat{\rho}(x)$, $\hat{\pi}(x)$ denote the estimated participation probability and propensity score models. We can see that extreme weights can result from small values of $\hat{\pi}(X_i)$, $1-\hat{\pi}(X_i)$ or $\hat{\rho}(X_i)$, which would occur when there is lack of overlap.
As a result, $ \hat{\theta}_\mathcal{T}$ will be highly influenced by a few observations of large weights and thus is highly sensitive to random errors in the observed outcomes. Other estimators based on probability weights, such as matching and stratification \citep{tipton2013improving}, also face the same issue.

Although not explicitly involving probability weights, methods based on outcome regression (OR) \citep{dahabreh2020extending}:
\begin{equation}
\label{eq:outcomeRegression}
    \hat{\theta}_{\mathcal{T}}^{(OR)} =
    \frac{\sum_{i=1}^n (1 - S_i)\{\hat{\mu}_{1,\mathcal{S}}(X_i) - \hat{\mu}_{0,\mathcal{S}}(X_i)\}}
         {\sum_{i=1}^n (1 - S_i)}
\end{equation}
are also vulnerable to the lack of overlap. Here $\hat{\mu}_{a, \mathcal{S}}(x), a\in\{0,1\}$ denote the outcome models fitted on the source sample. These methods essentially impute the potential outcomes for all the individuals in the target sample, thus requiring extrapolation on the covariate region with limited overlap. If $\hat{\mu}_{a, \mathcal{S}}(x)$'s are estimated with parametric models, such extrapolation is highly dependent on correct model specification, as we will further illustrate in Section \ref{sec:simu}. On the other hand, if $\hat{\mu}_{a, \mathcal{S}}(x)$'s are estimated with non-parametric models, such extrapolation is done based on a few observations in the source sample and thus is unreliable. 

Therefore, \cite{dahabreh2020extending} duly recommended examining the distribution of the estimated participation probabilities, or equivalently $\hat{\rho}(x)$ when the source sample is from a randomized trial with $\pi(x)=\pi$. However when the source sample is from an observational study, it is less clear how to summarize covariate overlap in terms of both $\rho(x)$ and $\pi(x)$. Furthermore, it is unknown how the other aspects about the potential outcomes in the source population would affect the estimation of $\theta_\mathcal{T}$.


\section{Methodology}
\label{sec:method}

\subsection{Efficiency bound} 
\label{sec:method-bound}

We consider restricting the estimation to a subpopulation of the target population. The subpopulation consists of individuals whose covariate values belong to a subset $B \subset \mathcal{X}$. Then we denote the ATE of the subpopulation as
\begin{equation}
	\label{eq:theta_B_T}
	\theta_{B, \mathcal{T}} = \E\{\theta(X)\mid X\in B, S = 0\}.
\end{equation}
In particular, $\theta_\mathcal{T} = \theta_{\mathcal{X}, \mathcal{T}}$. In what follows, we characterize the impact of overlap on the estimation precision of the aggregate causal effect $\theta_{B, \mathcal{T}}$ for any $B$. From our characterization, we propose a generalizability score to quantify the dissimilarity between an individual and the source population. For a fixed subset size, we can use the generalizability score to select subsets that are optimal for estimation efficiency.

To characterize the impact of overlap on the estimation of $\theta_{B, \mathcal{T}}$ without being tied to any specific estimator, our method is developed upon the semiparametric efficiency theory \citep{tsiatis2007semiparametric}. We focus on regular and asymptotically linear (RAL) estimators. The definition of RAL estimators can be found in \citet[Section 3.1]{tsiatis2007semiparametric}. Most commonly-used estimators are RAL, including all the estimators we will use in Sections \ref{sec:simu} and \ref{sec:data}. Let $\hat{\theta}_{B, \mathcal{T}}$ be an RAL estimator for $\theta_{B, \mathcal{T}}$, and we can write
\begin{equation}
\label{eq:decomp}
    \hat{\theta}_{B, \mathcal{T}} - \theta_{B, \mathcal{T}} = (\hat{\theta}_{B, \mathcal{T}} - \tilde{\theta}_{B, \mathcal{T}}) + (\tilde{\theta}_{B, \mathcal{T}} - \theta_{B, \mathcal{T}}).
\end{equation}
Here $\tilde{\theta}_{B, \mathcal{T}}$ is the sample version of \eqref{eq:theta_B_T}:
\begin{equation*}
    \tilde{\theta}_{B, \mathcal{T}} = 
    \frac
    {\sum_{i=1}^{n} (1-S_i)\ind_{B}(X_i)\theta(X_i)}
    {\sum_{i=1}^{n} (1-S_i)\ind_{B}(X_i) }
,\end{equation*}
where $\ind_{B}(X_i) = 1$ if $X_i \in B$ and 0 otherwise. The two terms on the right-hand side of \eqref{eq:decomp} are generally uncorrelated. Since the second term is the difference between a sample average of $\theta(X)$ and a population average, its uncertainty of the second term lies entirely on the sampling variability of $X$ and the treatment effect heterogeneity. Hence, the impact of insufficient overlap is fully captured by the first term. By focusing on the asymptotic variance of $\hat{\theta}_{B, \mathcal{T}} - \tilde{\theta}_{B, \mathcal{T}}$ instead of $\hat{\theta}_{B, \mathcal{T}} - \tilde{\theta}_{B, \mathcal{T}}$, we can better target our efforts on minimizing the impact of limited overlap. In contrast, taking the second term into consideration may result in a subset with little heterogeneity in order to minimize variance. Therefore, we state our theorem below on the variance bound of $\hat{\theta}_{B, \mathcal{T}} - \tilde{\theta}_{B, \mathcal{T}}$. 
\begin{theorem}
    \label{thm:eifB}
    Suppose Assumptions \ref{assp:unconf}-\ref{assp:ovlppop} hold, then for any RAL estimator $\hat{\theta}_{B, \mathcal{T}}$, the asymptotic variance of $\sqrt{n}(\hat{\theta}_{B, \mathcal{T}}-\tilde{\theta}_{B, \mathcal{T}})$ is at least
    \begin{equation}
        \label{eq:varB}
        \begin{aligned}
            V(B)
            &=
            \frac{1}{\E [\{1-\rho(X)\}\ind_{B}(X)]^2} 
            \E\left[
                \frac{\{1-\rho(X)\}^2\ind_{B}(X)}{\rho(X)} \left\{
                    \frac{\sigma^2_{1, \mathcal{S}}(X)}{\pi(X)} + \frac{\sigma^2_{0, \mathcal{S}}(X)}{1-\pi(X)}
                \right\}
            \right]
        .\end{aligned}
    \end{equation}
\end{theorem}

\medskip
This asymptotic variance bound holds regardless of whether the propensity score $\pi(x)$ is known or not. The participation probability $\rho(x)$ is taken as unknown as this is typically the case in practice. The proof of Theorem \ref{thm:eifB} is relegated to the Supplementary Materials. There we show that this bound is achieved by the semiparametric efficient estimator for $\theta_{B, \mathcal{T}}$.

\subsection{Generalizability score and generalization subset selection}
\label{sec:method-subsample}

Based on the expression of $V(B)$, we introduce a {\em generalizability score} 
\begin{equation}
	\label{eq:kappa}
	\kappa(x) = \frac{1-\rho(x)}{\rho(x)} \left\{
	\frac{\sigma^2_{1, \mathcal{S}}(x)}{\pi(x)} + \frac{\sigma^2_{0, \mathcal{S}}(x)}{1-\pi(x)}
	\right\}
\end{equation}
so that \eqref{eq:varB} can be expressed as
\begin{equation}
	\label{eq:varB2}
	V(B) = \frac{\E\left[ \kappa(X) \mid X \in B, S=0 \right]}{\prob(X \in B, S=0)}
.\end{equation}
The following theorem explores the relationship between $\kappa(x)$ and the set function $V(\cdot)$.

\begin{theorem}
\label{thm:optB}
Suppose Assumptions \ref{assp:unconf}-\ref{assp:ovlppop} hold. For any $\gamma > 0$, define
\begin{equation}
    \label{eq:B_gamma}
    B_\gamma = \left\{ 
        x\in \mathcal{X} 
        :
        \kappa(x)  
        \le \gamma
        \right\}
.\end{equation}
Then:
\begin{itemize}
	\item[(a)] For any $B \subset \mathcal{X}$ that satisfies $\prob(X \in B \mid S=0) = \prob(X \in B_\gamma \mid S=0)$, we have
	\begin{equation*}
	V(B_\gamma) \le V(B)
	.\end{equation*}
	\item[(b)] The optimal subset that minimizes $V(B)$ is $B^* = B_{\gamma^*}$, where $\gamma^* > 0$ satisfies
	\begin{equation}
		\label{eq:gamma_star}
		\gamma^* = 2 \E\left[ \kappa(X) \mid \kappa(X) \le \gamma^*, S=0 \right]
	.\end{equation}
\end{itemize}
\end{theorem}

\medskip
The proof of Theorem \ref{thm:optB} is relegated to the Supplementary Materials. Part (a) of the theorem indicates that $B_\gamma$ achieves the optimal efficiency bound among all the subsets that cover the same proportion of the target population. Since $B_\gamma$ is defined through $\kappa(x)$, this result suggests that we can use $\kappa(x)$ as a yardstick to rank and select subjects in the target population for generalization. The function $\kappa(x)$ effectively combines the participation probability $\rho(x)$ and the propensity score $\pi(x)$, unifying the two aspects of overlap into one single numerical value. 

Beside $\rho(x)$ and $\pi(x)$, $\kappa(x)$ also contains $\sigma^2_{1, \mathcal{S}}(x)$ and $\sigma^2_{0, \mathcal{S}}(x)$ which are the conditional variances of the potential outcomes in the source population. In theory, both $\sigma^2_{1, \mathcal{S}}(x)$ and $\sigma^2_{0, \mathcal{S}}(x)$  can be estimated from the source sample. However in practice, conditional quantities are hard to estimate precisely. One can either using smoothing techniques to roughly estimate these quantities or assume homoscedasticity and set $\sigma^2_{1, \mathcal{S}}(x) = \sigma^2_{0, \mathcal{S}}(x) =\sigma^2$ for some $\sigma^2$ \citep{crump2009dealing,li2018balancing,kallus2020more}. The exact value of $\sigma^2$ is not relevant to the use of $\kappa(x)$ because $\sigma^2$ is a constant multiplier and does not affect the relative scale of $\kappa(x)$. Therefore under the assumption of homoscedasticity, we can set $\sigma^2_{1, \mathcal{S}}(x) = \sigma^2_{0, \mathcal{S}}(x) =1$ in $\kappa(x)$. Then $\kappa(x)$ is completely determined by $\pi(x)$ and $\rho(x)$, and we don't need to use any observed outcome information in the source population when selecting the subpopulation for generalization. This eliminates the chance of introducing deliberate biases \citep{crump2006moving}. As highlighted in \citet{Rubinsim2007, rubin2008}, not peeking at the outcome information at the design phase is critical for assuring the objectivity of treatment effect estimation. For the rest of this article, we will set $\sigma^2_{1, \mathcal{S}}(x)= \sigma^2_{0, \mathcal{S}}(x) =  1$.

\begin{figure}[!p]
    \centering
    \includegraphics[width=.66\linewidth]{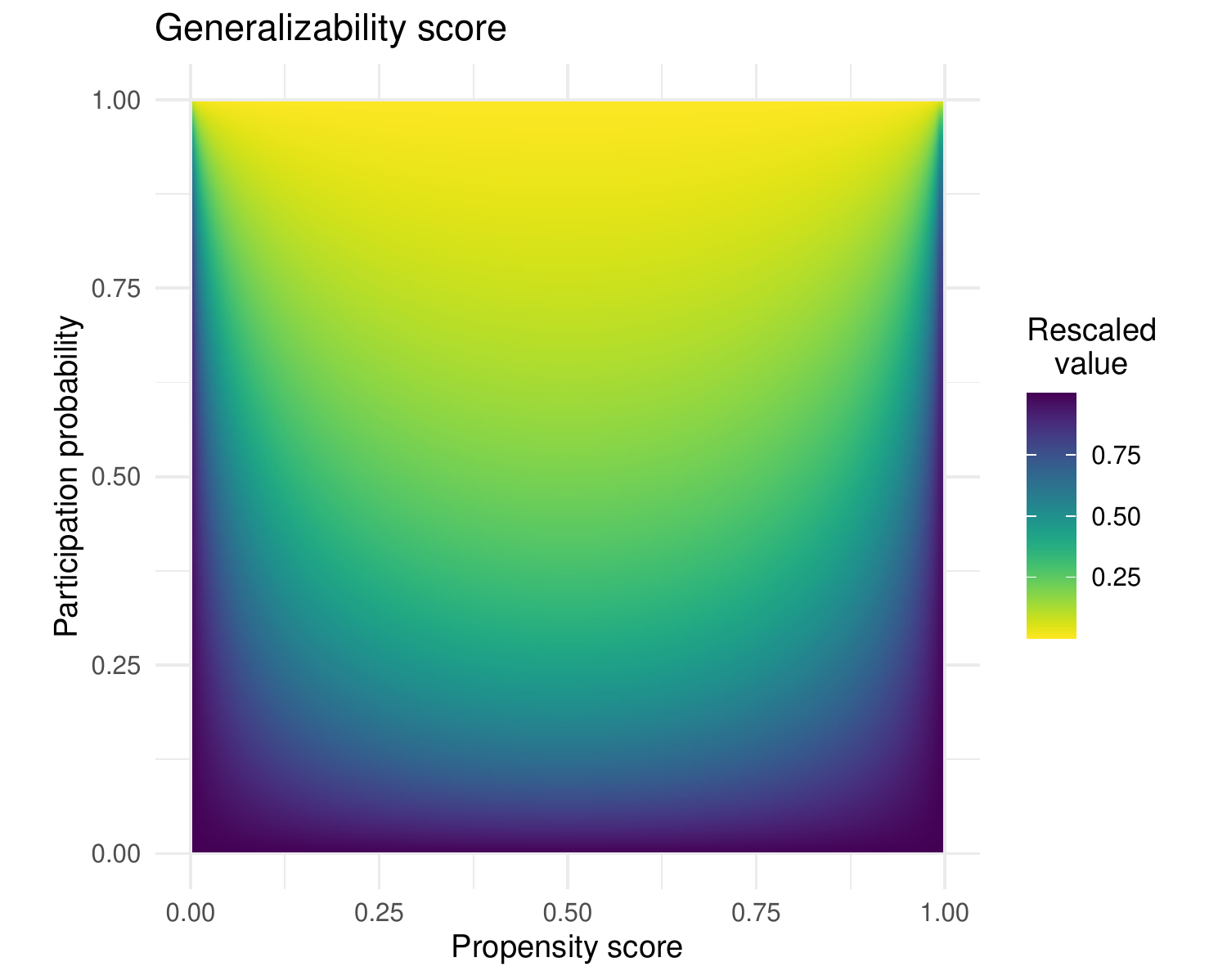}
    \caption{The generalizability score as a function of participation probability and propensity score under homoscedasticity. The generalizability score is rescaled onto $[0, 1]$ with the transformation $f(\kappa) = \kappa/(16 + \kappa)$. } 
    \label{fig:kappa}
\end{figure}

Figure \ref{fig:kappa} plots the generalizability score as a function  participation probability $\rho(x)$ and propensity score $\pi(x)$. Since $\kappa(x)$ changes rapidly near the margins of the plot region and has an unbounded range, we rescale it onto $[0,1]$ with a monotonic transformation, $f(\kappa) = \kappa / (16+\kappa)$, to help visualize it, where the choice of 16 is rather arbitrary. A high generalizability score is due to either a small $\rho(x)$, which suggests lack of overlap between the populations, or an extreme $\pi(x)$ close to 0 or 1, which indicates limited overlap between the treatment arms in the source population. So for individuals with a high generalizability score, it is harder to find close comparisons in the treated or control group in the source sample. That is why they are less suitable to be selected into the target subpopulation, as indicated by Theorem \ref{thm:optB}.

Choosing the size of $B_\gamma$, which is determined by the cut-off value $\gamma$, is also an important issue. On one hand, choosing a smaller cut-off will produce a subset with better overlap. On the other hand, a small cut-off will not only lead to a small target subsample, but may also have a reverse effect on the estimation precision since individuals in the source sample will also be excluded in estimation if they are outside the subset. Part (b) of Theorem \ref{thm:optB} provides a characterization of the optimal cut-off value that minimizes $V(B)$. The optimality condition \eqref{eq:gamma_star} is similar to a first-order optimality condition in common optimization problems. It does not guarantee a unique solution, though this is usually the case in our experience. However, it reveals an interesting characteristic of the optimal subset: the highest value of $\kappa(x)$ in the subset is no greater than twice its average, which suggests $\kappa(x)$ has a relatively even distribution on the subset. Therefore, any $\gamma$ that satisfies 
\begin{equation}
\label{eq:gamma_star_hat}
\gamma \le 2 \E \{\kappa(X) \mid \kappa(X)\le\gamma, S=0\}
\end{equation}
is a reasonable choice in the sense that the corresponding $B_\gamma$ does not contain individuals with extremely high $\kappa(x)$ as compared to the average level within the subset. Since the estimand $\theta_{B_\gamma, \mathcal{T}}$ is generally closer to $\theta_\mathcal{T}$ with a higher $\gamma$, in practice we can choose the cut-off value ${\gamma}^*$ to be the largest $\gamma$ that satisfies \eqref{eq:gamma_star_hat}. This is similar to the suggestion in \citet{crump2009dealing}. In particular, when $\kappa(X_i) \le 2 {\E}\{\kappa(X) \mid S=0\}$ for all $i \in \mathcal{T}$, the optimal target subsample would be the whole target sample. This subpopulation selection method will be used in our simulation studies and data analysis.

An alternative practical way to choose a desirable cut-off in practice is to compute $V({B}_\gamma)$ for some prespecified $\gamma$ values, such as some quantiles of $\{\kappa(X_i): i \in \mathcal{T}\}$, and then choose the largest $\gamma$ value that produces an acceptable $V({B}_\gamma)$. This method will be useful when one wishes to cover as much of the target population as possible in the generalizing aggregate causal effects without largely inflating the variance. This idea will be illustrated in Section \ref{sec:data}.


\section{Simulation Study}
\label{sec:simu} 

\subsection{Setup}
\label{sec:simu-setup}
\begin{figure}[!p]
	\centering
	\includegraphics[width=.99\linewidth]{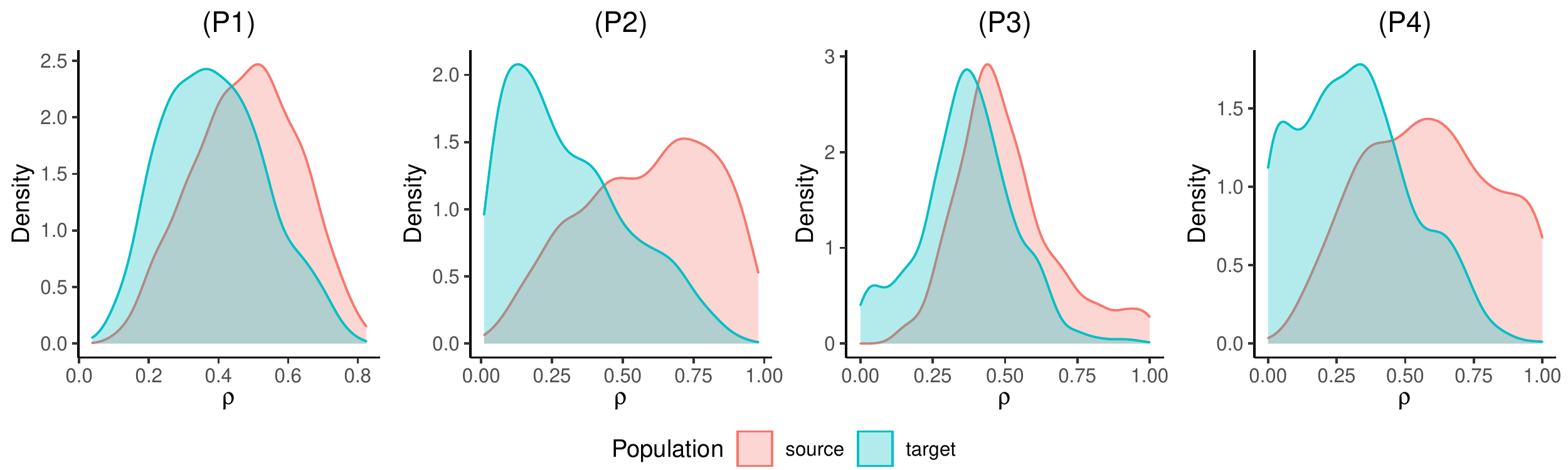}
	\caption{Overlap of participation probability in the simulation settings.}
	\label{fig:simu_rho_density}
\end{figure}

We adopt a rejection sampling procedure to generate covariates in both the source and target samples. In particular, we generate  $X_i = (X_{i1}, \dots, X_{i5})$ from a standard multivariate normal distribution $ N(\mathbf{0}_5, \mathbf{I}_5)$. Then we accept $X_i$ with probabilities $\tilde{\rho}(X_i)$ for the source sample (until sample sizes reach 600) and $1 - \tilde{\rho}(X_i)$ for the target sample  (until sample sizes reach 800), where $\tilde{\rho}(X_i)$ is specified as one of the following four models
\begin{itemize}
    \item[(P1)] $\tilde{\rho}(x) = \textnormal{logistic}(0.4 x_1 + 0.4 x_2 + 0.4 x_3)$,
    \item[(P2)] $\tilde{\rho}(x) = \textnormal{logistic}(0.8 x_1 + 0.8 x_2 + 0.8 x_3)$,
    \item[(P3)] $\tilde{\rho}(x) = \textnormal{logistic}(0.4 x_1 + 0.3 x_2^3 + 0.2 x_3^2)$,
    \item[(P4)] $\tilde{\rho}(x) = \textnormal{logistic}(0.8 x_1 + 0.6 x_2^3 + 0.4 x_3^2)$,
\end{itemize}
where $\textnormal{logistic}(z) = 1 / (1 + e^{-z})$. Under this sampling mechanism we have the density of the source covariates $p_s(x) \propto p_{\textnormal{normal}}(x)\tilde{\rho}(x)$ and of the target covariates $p_t(x) \propto p_{\textnormal{normal}}(x)\{1-\tilde{\rho}(x)\}$, where $p_{\textnormal{normal}}(.)$ is the density of a standard normal distribution. Thus the true participation probability model is $\rho(x) = \textnormal{logistic}(c + \textnormal{logit}(\tilde{\rho}(x)))$ for some constant $c$, where $\textnormal{logit}(z) = \log(z/(1-z))$ is the inverse of the logistic function. Figure \ref{fig:simu_rho_density} plots the distributions of the participation probability for the source and target populations under each of these settings. As can be seen, in (P1) and (P3) the source and target samples have relatively good overlap, whereas in (P2) and (P4) they have relatively bad overlap.

In the source sample, we set $\pi(x) = \textnormal{logistic}(0.3 x_1 - 0.3 x_3)$ and simulate the treatment assignments by $A_i \sim \textnormal{Bernoulli}(\pi(X_i))$. The observed outcomes are generated as $Y_i = (1 - A_i) \mu_{0, \mathcal{S}}(X_i) + A_i \mu_{1, \mathcal{S}}(X_i) + N(0, 1)$. Both linear and nonlinear potential outcome models ($a=0, 1$) are considered:
\begin{itemize}
    \item[(O1)] $\mu_{a, \mathcal{S}}(x) = x_1 + a(0.5x_1 + x_2 + 1)$,
    \item[(O2)] $\mu_{a, \mathcal{S}}(x) = x_1 - x_4 + (a - 0.5) \{0.4 (x_1 - 0.5)^2 + 0.5 x_2^2\}$.
\end{itemize}

To carry out the estimation of $\theta_{B, \mathcal{T}}$ we consider estimators of the following form, which can be found in \citet{dahabreh2020extending}:
\begin{equation}
    \label{eq:estimatorform}
    \begin{aligned}
        \hat{\theta}_{B, \mathcal{T}} 
        &= 
        \frac{\sum_{i:X_i\in B} w_{i1} S_i A_i  (Y_i - u_{i1})}
             {\sum_{i:X_i\in B} w_{i1} S_i A_i } 
		-
        \frac{\sum_{i:X_i\in B} w_{i0} S_i (1-A_i) (Y_i - u_{i0})}
             {\sum_{i:X_i\in B} w_{i0} S_i (1-A_i)} 
        \\&\quad+
        \frac{\sum_{i:X_i\in B} (1-S_i)(u_{i1} - u_{i0})}
             {\sum_{i:X_i\in B} (1-S_i)}
    .\end{aligned}
\end{equation}
When $w_{ia} = \hat{w}_{ia}$ as defined in \eqref{eq:weights} and $u_{ia} = 0$, $\hat{\theta}_{B, \mathcal{T}}$ corresponds to the IPW estimator \eqref{eq:Hajek}, but is restricted on the subset $B$. When $w_{ia} = 0$ (with the first two terms of \eqref{eq:estimatorform} set as 0) and $u_{ia} = \hat{\mu}_{a,\mathcal{S}}(X_i)$, it corresponds to the OR estimator \eqref{eq:outcomeRegression}. Moreover, when $w_{ia} = \hat{w}_{ia}$ and $u_{ia} = \hat{\mu}_{a,\mathcal{S}}(X_i)$, this gives rise to the augmented IPW (AIPW) estimator, which can yield consistent estimates under correct specification of either the propensity score and participation probability models or of the outcome regression models.

In our implementation, both $\pi(x)$ and $\rho(x)$ are estimated with logistic regression models and the outcome models are estimated with simple linear regression models. The true propensity score model is correctly specified under all scenarios. In the Supplementary Materials we include additional results under misspecified propensity score models. The participation probability model is correctly specified under (P1) and (P2) but misspecified under (P3) and (P4), and the outcome models are only correctly specified under (O1).

The probability estimates $\hat{\pi}(x)$ and $\hat{\rho}(x)$ from the full sample are used to compute the simplified version of the generalizability score, denoted by $\hat{\kappa}(x)$, which is based on \eqref{eq:kappa} but with $\sigma^2_{a, \mathcal{S}}(x)$ set to 1 for $a \in \{0, 1\}$. Then $\hat{\kappa}(x)$ is used to construct $\widehat{B}^* = \{x \in \mathcal{X}: \hat{\kappa}(x) \le \hat{\gamma}^*\}$, where $\hat{\gamma}^*$ is the largest $\gamma$ that satisfies
$$
	\gamma \le 2 \hat{\E} \{\hat{\kappa}(X) \mid \hat{\kappa}(X)\le\gamma, S=0\}
$$
according to \eqref{eq:gamma_star_hat} and $\hat{\E}(\cdot)$ denotes the empirical mean. Once $\widehat{B}^*$ is selected, all the probabilities and outcome models are re-estimated using the subsample within $\widehat{B}^*$ before computing the $\hat{\theta}_{\widehat{B}^*, \mathcal{T}}$, as advocated by \citet{crump2009dealing,li2019addressing}.

\subsection{Results}
\label{sec:simu-result}
We study the estimation precision improvement from subpopulation selection by contrasting the estimation results for the full target population to those for the subpopulation in $\widehat{B}^*$, which are measured in terms of bias, root mean square error (RMSE), and the average width and coverage rate of 95\% confidence interval (CI). Note that the subpopulation results are with respect to $\theta_{\widehat{B}^*, \mathcal{T}}$, the ATE on the selected subpopulation.
To construct the 95\% CIs for each estimator, in each simulation run we estimate the standard error using the bootstrap \citep{efron1994introduction}, and then set the CI as the range within 1.96 standard error of the corresponding point estimate.

Table \ref{tab:simu} summarizes the results based on 1000 repetitions. We also report the average proportion of the target sample that is kept in the subset $\widehat{B}^*$ under each scenario. We can see that for all three estimators, restricting the estimation to $\widehat{B}^*$ results in smaller RMSE and narrower CIs, especially when the overlap is insufficient ((P2) and (P4)). The IPW estimator gains its efficiency because the estimation weights $w_{ia}$ are more stable and less likely to have extreme values over $\widehat{B}^*$; the OR estimator gets improved because the outcome models have a better fit on the target sample when restricted on the region $\widehat{B}^*$ with good overlap; the AIPW estimator benefits from both aspects. When there is model misspecification, the bootstrap CIs are more likely to achieve the nominal coverage rate when restricting the analysis on the subpopulation. Under (O1), we observe less improvement for the OR and AIPW estimators; this is because when a parametric outcome model is correctly specified, the issue of lacking overlap is largely mitigated by model extrapolation. Additional simulation results for studying impact of incorrectly specified the propensity score model and the impact of using a larger cut-off value can be found in Section S.4 of the Supplementary Materials. Under these settings, we also observe substantial precision improvement for the estimation from restricting to a subpopulation.


\section{Coordinated-Transitional Care (C-TraC) Program}
\label{sec:data}
In this section, we illustrate the proposed approach by evaluating the treatment effect of C-TraC Program versus the standard care on 30-day rehospitalization \citep{gilmore2014development}. The C-TraC Program is a telephone-based, protocol-driven intervention designed to support and empower patients to properly manage their post-discharge care. The population of interest for this program is mainly patients who are 65 years or older (Medicare patients) and who have been hospitalized. Due to both patient factors such as limited cognition or living alone and system factors such as lack of adequate transitional care and lack of patient education, such a population has a high tendency to be readmitted to a hospital within 30 days of being discharged. 

The observational study consists of patients who met the inclusion criteria and were discharged from the UW-Hospital between January 2013 and April 2018. We break this data set in the middle and consider patients who were discharged prior to 2016 as the source sample, and the rest as the target sample. In this way, we also can compare our estimated target sample causal quantities with the actual observed outcomes in the target sample. 

Among the source sample, 206 patients participated in the C-TraC program, among which 42 were readmitted to the hospital within 30 days, and 507 patients didn't participate in the program, among which 109 were rehospitalized. Program participation and rehospitalization information is also available for the target sample, but will be held out from the estimation and used as a benchmark to evaluate the estimation accuracy.

With the input from our collaborators, we use ten covariates measured at study entry date as covariates. They include two risk scores. The well-known LACE index score for risk of readmission/death within thirty days of discharge \citep{vanWalraven551} and Hendrich II Fall Risk score for high risk patient falls identification \citep{HENDRICH20039}. The LACE score is based on four features of an inpatient hospital episode: length of stay (LoS), admission type, comorbidities and the number of accident and emergency (A\&E) visits made by a patient in the 6 months prior to their initial admission.  The Hendrich score is based on 7 risk factors in its model. Besides these two scores, the other variables include age, disease status (diabetes, cancer history, respiratory symptoms, and malnutrition), prescription medication for malnutrition, and lab test markers for malnutrition/liver/kidney functions (ALB) and liver disease (ALT).

We estimate the propensity score $\pi(x)$ by applying logistic regression on the source sample, and estimate the participation probability $\rho(x)$ by applying logistic regression on the combined sample. 
To assess the fit of the probability estimates, we check the covariate balances by performing (unweighted and weighted) one-way ANOVA on each of the covariates across three groups: treated, control and target sample. The results are summarized in the 2nd to 5th columns of Table \ref{tab:covariate_Bal}, where ``weighted'' means inverse probability weights are used in the fitting the ANOVA. After weighting, the balance of all the covariates across the three groups is largely improved. This suggests that the logistic models provide a reasonable fit for $\pi(x)$ and $\rho(x)$.

\begin{figure}[!p]
    \centering
    \includegraphics[width=.66\linewidth]{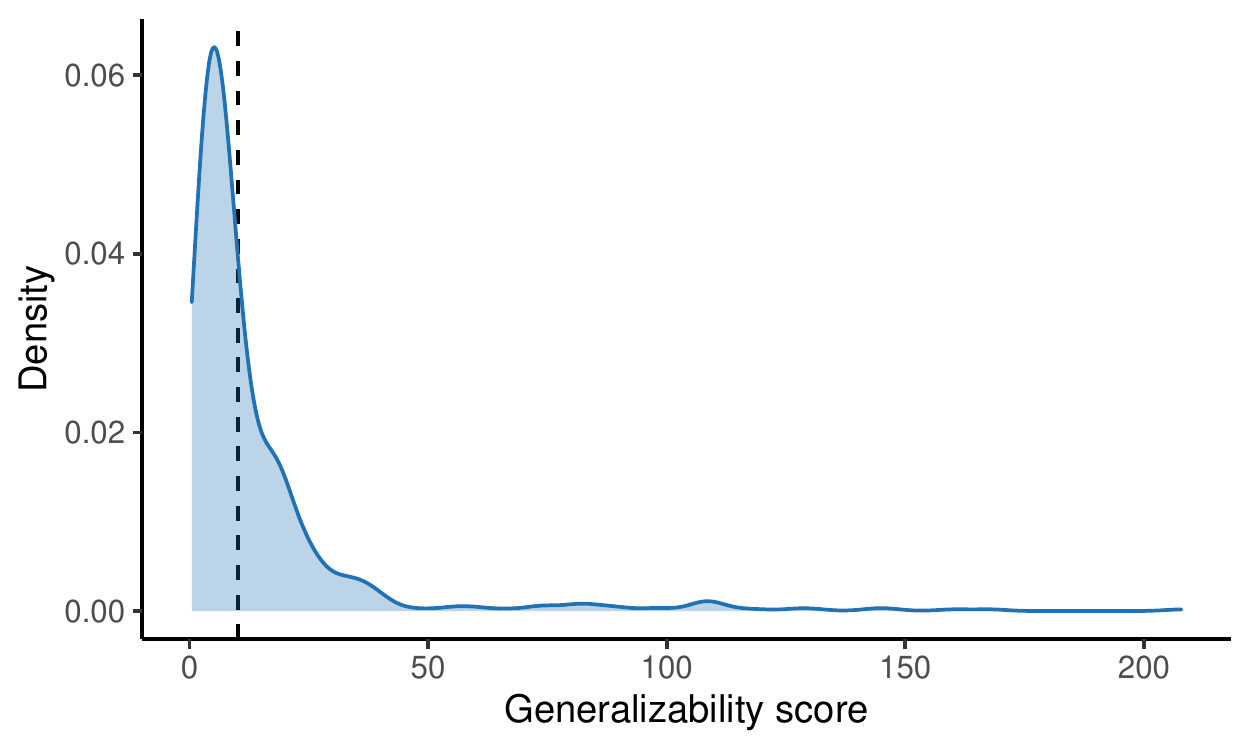}
    \caption{Distribution of the generalizability score among the target sample. The vertical dashed line corresponds to the estimated optimal cut-off $\hat{\gamma}^*$.}
    \label{fig:data_kappa}
\end{figure}

We compute the generalizability score, denoted by $\hat{\kappa}(x)$, using the fitted participation probability and propensity score models and substituting $\sigma^2_{a, \mathcal{S}}(x)$ with 1. We show the distribution of the generalizability score among the target sample in Figure \ref{fig:data_kappa}. The distribution has a long tail on the right, which suggests insufficient overlap for these individuals. The vertical dashed line represents $\hat{\gamma}^*$ the largest value of $\gamma$ that satisfies \eqref{eq:gamma_star_hat}, which is used to construct the subset $\widehat{B}^* = \{x \in \mathcal{X}: \hat{\kappa}(x) \le \hat{\gamma}^*\}$. About 64.1\% of the target sample and 84.3\% of the source sample are included in this subset. The last two columns of Table \ref{tab:covariate_Bal} report the covariate imbalance for the subsample after weighting adjustment, which confirm that the logistic models remain a good fit for $\pi(x)$ and $\rho(x)$.

We estimate the ATE and its standard error on the whole covariate region $\mathcal{X}$ and on the selected subset $\widehat{B}^*$, respectively. The estimators are the same as those given by \eqref{eq:estimatorform}. Since we are dealing with binary outcomes, for the OR and AIPW approaches, outcome regression models are fitted using logistic regression. Standard error estimates are obtained from 2000 bootstrap replications. These estimates with the standard errors are displayed in rows 1 and 3 of Table \ref{tab:data_estimate}. To benchmark the accuracy of the estimates, we also conduct standard ATE estimation using the treatment assignment and outcome information from the target sample, and display these estimates in rows 2 and 4 of the table. Such standard estimates would be closer to the true values, but are infeasible in the generalization setting. For any given method, focusing on $\widehat{B}^*$ (rows 3 and 4) leads to much closer estimates to the infeasible standard estimates, compared to the results on the whole covariate region (rows 1 and 2).

\begin{figure}[!p]
    \centering
    \includegraphics[width=.95\linewidth]{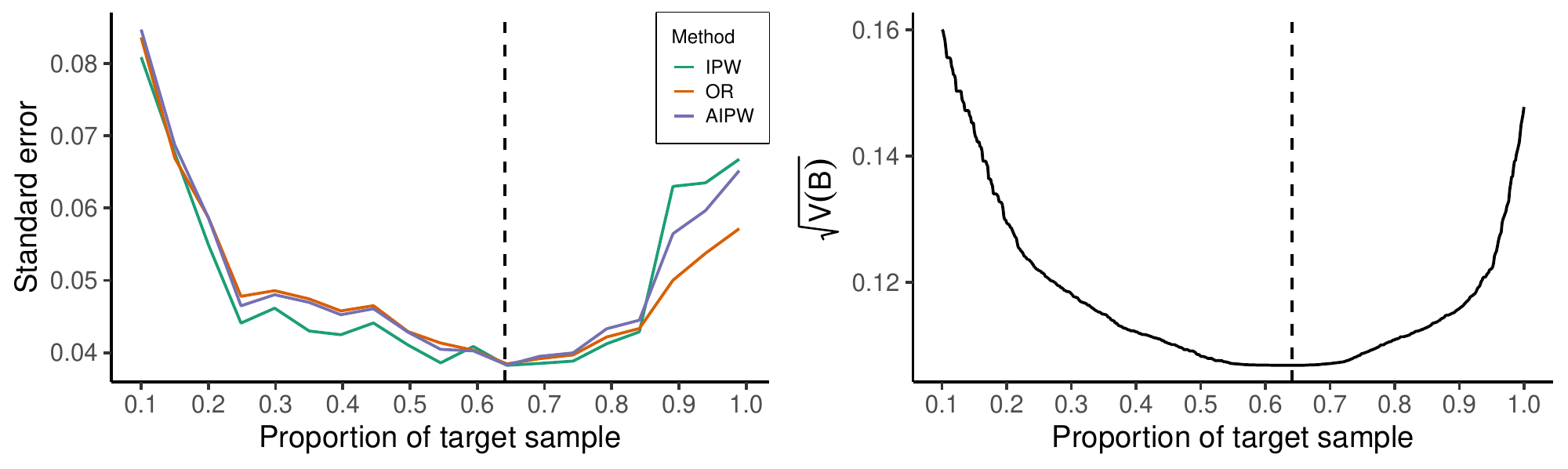}
    \caption{Left panel: bootstrap standard error estimates. Right panel: $V(\widehat{B}_\gamma)^{\nicefrac{1}{2}}$. The dashed vertical line represents the proportion contained in $\widehat{B}^*$.}
    \label{fig:var_thr}
\end{figure}

To further investigate the impact of using other cut-off values in the subpopulation selection process, we repeat the aforementioned procedure of estimating the treatment effect and standard error for subsets of the form $\widehat{B}_\gamma = \{x \in \mathcal{X}: \hat{\kappa}(x) \le \gamma\}$, with $\gamma$ being the 10th, 15th, \dots, 100th percentiles of $\{\hat{\kappa}(X_i): i \in \mathcal{T}\}$. Accordingly, the subsamples under these cut-offs cover 10\%, 15\%, \dots, 100\% of the target sample, respectively. The left panel of Figure \ref{fig:var_thr} plots the bootstrap standard error estimates against the proportion of the target sample covered. The vertical dashed line corresponds to $\hat{\gamma}^*$. As we can see, the curves for the three estimators follow a very similar pattern: as the size of the subpopulation grows, the standard errors first decrease as the sample size increases, but then bounce back as the impact of limited overlap dominates. All the curves attain the minimum at around the proportion corresponding to $\hat{\gamma}^*$. It is worth noticing that all the curves is relatively flat around the minimum. Hence, if we select a cut-off value slightly higher than $\hat{\gamma}^*$, we can obtain a larger subpopulation without heavily sacrificing the estimation precision. This might be preferable in practice because with a larger subpopulation the corresponding estimand is generally closer to the ATE on the entire target population.

For comparison, we also evaluate $V(\widehat{B}_\gamma)$ plugging $\kappa(x)$ into \eqref{eq:varB2}. This is done for all $\gamma \in \{\hat{\kappa}(X_i): i \in \mathcal{T}\}$, each cut-off corresponding to a subsample of different size. On the right panel of Figure \ref{fig:var_thr} we plot $V(\widehat{B}_\gamma)^{\nicefrac{1}{2}}$ against the proportion of target sample that $\widehat{B}_\gamma$ contains. Although $V(\widehat{B}_\gamma)$ is based on efficiency bound rather than the actual variance of the given estimators and is computed with the homoscedasticity simplification, its value follows a very similar trend to the standard error curves on the left panel. The computation of $V(\widehat{B}_\gamma)$ is much more efficient than bootstrap resampling and, more importantly, does not require using any outcome information. Therefore, we can also use $V(\widehat{B}_\gamma)$ to guide our selection of $\gamma$.

Lastly, the variance bound $V(B)$ in \eqref{eq:varB2} is derived regardless of how $B$ is selected, and is a metric of the impact of overlap for any subset. Our subset selection has focused on the sublevel sets of the generalizability score $\kappa(x)$, which could be complex if the functional form of $\kappa(x)$ is complex. If one wishes to construct a more regular subset that is easier to describe, we can apply the tree building procedure in \citet{traskin2011defining} using $V(B)$ as a guidance. Specifically, suppose we have constructed a subset $\widehat{B}$ by trimming $\kappa(x)$, and we label the units as $1$ if they are within $\widehat{B}$ and $0$ otherwise, then we can build a classification tree to approximate these labels. Let $\widetilde{B}$ be the subset given by the classification tree. The ratio $ V(\widetilde{B}) / V(\widehat{B}) $ can be used as a guidance for tuning the complexity of the tree. The resulting $\widehat{B}$ would be easier to interpret as it may depend on a few covariates. In the Supplementary Materials we provide an illustration of this tree building procedure applied to C-TraC Program study.


\section{Discussion}
\label{sec:disc}

In this paper, we systematically study the impact of covariate overlap on generalizing ATE estimation to a target population. In order to deal with the issue of insufficient overlap that one often encounters in real applications, we propose to limit our attention to subpopulations of the target population. The resulting ATE estimand might differ from the ATE on the entire target population, but allows for a much more stable and reliable estimation. We quantify the impact of overlap based on the semiparametric efficiency theory, and derive a generalizability score to guide the selection of subpopulations. The generalizability score summarize the variance inflation due to insufficient overlap in both the propensity score and participation probability simultaneously. To reduce estimation variance due to insufficient overlaps, individuals with a large generalizability score are less suitable to be included in the subpopulation. We also empirically demonstrate that one could assume homoscedasticity of the potential outcomes and set the conditional variance terms as constant in the generalizability score. In this way, we can select the subpopulation without peeking at the observed outcomes from the source population, thus avoiding deliberate biases. The simulation and real data analysis results demonstrate substantial precision improvement by utilizing the generalizability score.

When applying our method, practitioners should be aware that the resulting ATE estimation is with respect to the subpopulation selected, which is defined by trimming individuals with anomalously high generalizability score. We have characterized the optimal cut-off value $\gamma^*$ that minimizes the impact of limited overlap; however, it might be desirable in practice to use a cut-off moderately higher than $\gamma^*$ to cover a larger proportion of the target population. As demonstrated in Section \ref{sec:data}, this may not severely compromise the estimation precision. A graphical approach (based on the right panel of Figure \ref{fig:var_thr}) might be a practical alternative method to select the cut-off.

Throughout, we have focused on the scenarios where the source sample is external to the target population ($\mathcal{S} \not \subset \mathcal{T}$), which is also known as non-nested design.
In the setting of nested design ($\mathcal{S} \subset \mathcal{T}$), for example, generalizing the result from a randomized trial to all trial-eligible individuals \citep{Cole2010}, 
our proposed approach can be easily adapted to address similar overlap issues. In this case, the efficiency bound would become
$$
\E \left[
    \ind_{B}(X) \left\{ 
        \frac{\sigma^2_{1, \mathcal{S}}(X)}{\rho(x)\pi(x)} + \frac{\sigma^2_{0, \mathcal{S}}(X)}{\rho(x)(1 - \pi(x))}
    \right\}  
\right] \bigg / \prob(B)^2,
$$ 
and we can similarly define the generalizability score as $\tilde{\kappa}(x) = \sigma^2_{1, \mathcal{S}}(X)/\{\rho(x)\pi(x)\} + \sigma^2_{0, \mathcal{S}}(X)/[\rho(x)\{1 - \pi(x)\}]$.

\section{Supplementary Materials}

The codes for this paper are available at \url{https://github.com/DRuiCHEN/genScore}.

 \section*{Acknowledgments}
 We thank the reviewers, the associate editor and co-editors for their helpful comments that greatly improved this paper.
 Research reported in this work was funded through a Patient-Centered Outcomes Research
 Institute (PCORI) Award (ME-2018C2-13180). The views in this work are solely the responsibility
 of the authors and do not necessarily represent the views of the Patient-Centered
 Outcomes Research Institute (PCORI), its Board of Governors or Methodology Committee.\\
 {\it Conflict of Interest}: None declared.

\bibliographystyle{biorefs}
\bibliography{ref_generalizability}

\begin{table*}[!p]
    \small
    \centering
    \caption{Simulation results (the biases and RMSEs are multiplied by 10).}
    \label{tab:simu}
    \begin{tabular}{
        c
        SSSS[table-format = 2.1]<{\si{\percent}}
        SSSS[table-format = 2.1]<{\si{\percent}}
		c
    }
    \toprule
    \multirow{3}{*}{Method} & 
    \multicolumn{4}{c}{Full target population} &
    \multicolumn{4}{c}{Optimal subpopulation in $\widehat{B}^*$} &
    \multirow{3}{*}{\begin{tabular}{@{}c@{}} Subpopulation \\ proportion$^1$ \end{tabular}} 
    \vspace{-1.5pt} \\ 
	\cmidrule(lr){2-5}\cmidrule(lr){6-9} \vspace{-13pt}\\ \vspace{-1pt}
    & 
	\multicolumn{1}{c}{Bias} & \multicolumn{1}{c}{RMSE} & \multicolumn{1}{c}{CI} & \multicolumn{1}{c}{CI} &
	\multicolumn{1}{c}{Bias} & \multicolumn{1}{c}{RMSE} & \multicolumn{1}{c}{CI} & \multicolumn{1}{c}{CI} &
	\\
	& 
	\multicolumn{1}{c}{($\times 10$)} & \multicolumn{1}{c}{($\times 10$)} & \multicolumn{1}{c}{width} & \multicolumn{1}{c}{coverage} & 
	\multicolumn{1}{c}{($\times 10$)} & \multicolumn{1}{c}{($\times 10$)} & \multicolumn{1}{c}{width} & \multicolumn{1}{c}{coverage} & 
	\\
    \midrule 
    \vspace{-8pt} \\ 
	\multicolumn{10}{l}{\small \hspace{7.5em} (O1)(P1) \textit{outcome model correct, $\rho$ model correct, good overlap}} \vspace{3pt} \\
	IPW  & 0.03  & 1.69  & 0.65  & 94.5  & 0.06  & 1.23  & 0.50  & 94.9  & \multirow{3}{*}{80.6\%} \\
	OR  & 0.05  & 1.06  & 0.42  & 95.4  & 0.07  & 1.02  & 0.41  & 96.1  & \\
	AIPW  & 0.05  & 1.12  & 0.44  & 95.5  & 0.08  & 1.04  & 0.42  & 95.4  & \vspace{4pt}\\
	\multicolumn{10}{l}{\small \hspace{7.5em} (O1)(P2) \textit{outcome model correct, $\rho$ model correct, bad overlap}} \vspace{3pt} \\
	IPW  & 0.15  & 3.49  & 1.10  & 91.6  & 0.01  & 1.44  & 0.59  & 95.8  & \multirow{3}{*}{44.2\%} \\
	OR  & 0.03  & 1.35  & 0.53  & 95.4  & 0.00  & 1.16  & 0.47  & 96.0  & \\
	AIPW  & -0.02  & 1.74  & 0.64  & 93.5  & 0.01  & 1.21  & 0.49  & 95.9  & \vspace{4pt}\\
	\multicolumn{10}{l}{\small \hspace{7.5em} (O1)(P3) \textit{outcome model correct, $\rho$ model incorrect, good overlap}} \vspace{3pt} \\
	IPW  & 0.51  & 1.76  & 0.63  & 92.2  & -0.17  & 1.26  & 0.48  & 93.5  & \multirow{3}{*}{78.1\%} \\
	OR  & -0.07  & 1.11  & 0.43  & 94.5  & -0.00  & 1.00  & 0.40  & 94.2  & \\
	AIPW  & -0.08  & 1.16  & 0.44  & 93.2  & 0.00  & 1.03  & 0.40  & 95.0  & \vspace{4pt}\\
	\multicolumn{10}{l}{\small \hspace{7.5em} (O1)(P4) \textit{outcome model correct, $\rho$ model incorrect, bad overlap}} \vspace{3pt} \\
	IPW  & 0.99  & 2.75  & 0.92  & 89.0  & -0.11  & 1.35  & 0.53  & 95.1  & \multirow{3}{*}{52.0\%} \\
	OR  & 0.03  & 1.23  & 0.50  & 96.2  & 0.05  & 1.10  & 0.44  & 96.1  & \\
	AIPW  & 0.03  & 1.46  & 0.56  & 94.8  & 0.05  & 1.15  & 0.46  & 95.5  & \vspace{4pt}\\
	\multicolumn{10}{l}{\small \hspace{7.5em} (O2)(P1) \textit{outcome model incorrect, $\rho$ model correct, good overlap}} \vspace{3pt} \\
	IPW  & -0.03  & 1.63  & 0.61  & 94.2  & -0.03  & 1.25  & 0.50  & 96.1  & \multirow{3}{*}{80.9\%} \\
	OR  & -1.40  & 1.82  & 0.44  & 76.0  & -0.61  & 1.22  & 0.42  & 91.6  & \\
	AIPW  & -0.07  & 1.32  & 0.48  & 93.1  & -0.02  & 1.09  & 0.42  & 94.4  & \vspace{4pt}\\
	\multicolumn{10}{l}{\small \hspace{7.5em} (O2)(P2) \textit{outcome model incorrect, $\rho$ model correct, bad overlap}} \vspace{3pt} \\
	IPW  & -0.10  & 3.06  & 1.05  & 92.8  & -0.05  & 1.59  & 0.62  & 95.1  & \multirow{3}{*}{44.4\%} \\
	OR  & -3.48  & 3.78  & 0.57  & 33.0  & -0.79  & 1.47  & 0.47  & 88.9  & \\
	AIPW  & -0.17  & 2.32  & 0.77  & 90.1  & -0.02  & 1.32  & 0.49  & 92.6  & \vspace{4pt}\\
	\multicolumn{10}{l}{\small \hspace{7.5em} (O2)(P3) \textit{outcome model incorrect, $\rho$ model incorrect, good overlap}} \vspace{3pt} \\
	IPW  & -1.14  & 1.85  & 0.59  & 87.9  & 0.09  & 1.21  & 0.49  & 95.8  & \multirow{3}{*}{77.9\%} \\
	OR  & -3.07  & 3.26  & 0.44  & 21.5  & -0.55  & 1.15  & 0.41  & 92.6  & \\
	AIPW  & -1.41  & 1.85  & 0.48  & 78.5  & 0.17  & 1.04  & 0.41  & 95.1  & \vspace{4pt}\\
	\multicolumn{10}{l}{\small \hspace{7.5em} (O2)(P4) \textit{outcome model incorrect, $\rho$ model incorrect, bad overlap}} \vspace{3pt} \\
	IPW  & -1.62  & 2.82  & 0.86  & 86.0  & 0.41  & 1.49  & 0.60  & 95.1  & \multirow{3}{*}{51.6\%} \\
	OR  & -5.43  & 5.60  & 0.53  & 2.9  & -0.40  & 1.24  & 0.46  & 93.6  & \\
	AIPW  & -2.33  & 2.89  & 0.63  & 62.3  & 0.53  & 1.32  & 0.47  & 92.6  & \\
  	\bottomrule
    \end{tabular}
\begin{tablenotes}
  \item $^1$Average proportion of the target sample that are kept in $\widehat{B}^*$.
\end{tablenotes}
\end{table*}

\begin{table}[!p]
    \centering
    \small
    \caption{Covariate balance checking.}
    \label{tab:covariate_Bal}
    \begin{tabular}{
        l@{\hskip 10pt}
        S@{\hspace{3pt}}S@{\hspace{12pt}}
        S@{\hspace{3pt}}S@{\hspace{12pt}}
        S@{\hspace{3pt}}S
    }
    \toprule
    \multicolumn{1}{c}{\multirow{2}{*}{Variable}}     & 
    \multicolumn{2}{c}{unweighted}  & 
    \multicolumn{2}{c}{weighted}  & 
    \multicolumn{2}{c}{subset, weighted} 
    \vspace{-1pt}\\ \cmidrule(lr){2-3}\cmidrule(lr){4-5}\cmidrule(lr){6-7} \vspace{-10pt} \\
    \multicolumn{1}{c}{} & 
    \multicolumn{1}{c}{F-statistic} & \multicolumn{1}{c}{p-value} & 
    \multicolumn{1}{c}{F-statistic} & \multicolumn{1}{c}{p-value} & 
    \multicolumn{1}{c}{F-statistic} & \multicolumn{1}{c}{p-value} \\
    \midrule
	{Age}  & 8.89  & 0.000  & 0.39  & 0.676  & 0.06  & 0.941  \\
	{Diabetes w/ chronic complications}  & 5.00  & 0.007  & 1.56  & 0.210  & 0.06  & 0.946  \\
	{Breast, Lung, and Other Cancers}  & 15.70  & 0.000  & 0.88  & 0.413  & 0.05  & 0.951  \\
	{Respiratory symptoms}  & 8.82  & 0.000  & 0.10  & 0.906  & 0.01  & 0.993  \\
	{Malnutrition}  & 11.06  & 0.000  & 0.85  & 0.429  & 0.21  & 0.813  \\
	{Therapeutic Nutrients (Rx)}  & 17.95  & 0.000  & 1.73  & 0.178  & 0.11  & 0.892  \\
	{Lab: Albumin (ALB)}  & 13.02  & 0.000  & 0.23  & 0.793  & 0.24  & 0.788  \\
	{Lab: Alanine Amino Transferase (ALT)}  & 19.29  & 0.000  & 0.07  & 0.933  & 0.24  & 0.786  \\
	{LACE Index}  & 18.33  & 0.000  & 0.04  & 0.961  & 0.07  & 0.937  \\
	{Hendrich II Fall Risk}  & 27.49  & 0.000  & 0.25  & 0.775  & 0.01  & 0.991  \\
    \bottomrule
    \end{tabular}
\end{table}

\begin{table}[!p]
    \centering
    \small
    \caption{ATE estimates and standard errors (all numbers are multiplied by 100).}
    \label{tab:data_estimate}
    \begin{tabular}{cc|ScScSc}
    \toprule
    \multirow{2}{*}{Estimand} & 
    \multirow{2}{*}{\begin{tabular}{@{}c@{}} Generalized \\ from $\mathcal{S}$? \end{tabular}} & 
	\multicolumn{2}{c}{\multirow{2}{*}{IPW}} & 
	\multicolumn{2}{c}{\multirow{2}{*}{OR}} & 
	\multicolumn{2}{c}{\multirow{2}{*}{AIPW}} 
	\\ &&& \\
    \midrule 
    \vspace{-10pt}\\
	\multirow{2}{*}{$\theta_\mathcal{T}$}
	& Yes   & -1.41  & \hspace{-14pt}(6.80) \hspace{-7pt}  & -3.04  & \hspace{-14pt}(6.00) \hspace{-7pt}  & -1.11  & \hspace{-14pt}(6.72) \hspace{-7pt}  \\
	& No$^1$  & -7.51  & \hspace{-14pt}(3.14) \hspace{-7pt}  & -7.57  & \hspace{-14pt}(3.12) \hspace{-7pt}  & -7.39  & \hspace{-14pt}(3.15) \hspace{-7pt}  \\
	\cmidrule{1-8}
	\multirow{2}{*}{$\theta_{\widehat{B}^*, \mathcal{T}}$}
	& Yes  & -2.21  & \hspace{-14pt}(3.88) \hspace{-7pt}  & -1.45  & \hspace{-14pt}(3.88) \hspace{-7pt}  & -1.90  & \hspace{-14pt}(3.88) \hspace{-7pt}  \\
	& No  & -1.41  & \hspace{-14pt}(3.66) \hspace{-7pt}  & -1.85  & \hspace{-14pt}(3.61) \hspace{-7pt}  & -1.73  & \hspace{-14pt}(3.65) \hspace{-7pt}  \\
	\bottomrule
    \end{tabular}
    \begin{tablenotes}
		\item \hspace{3em}$^1$Estimated using outcome information in the target sample, used as comparison benchmark.
    \end{tablenotes}
\end{table}

\end{document}